\documentclass[onecolumn]{article}    
\usepackage{xcolor}
\usepackage{cite}
\usepackage{amsmath,amssymb,amsfonts}
\usepackage{amsopn} 
\usepackage{algorithmic}
\usepackage{graphicx}
\usepackage{algorithm,algorithmic}
\usepackage{hyperref}
\hypersetup{hidelinks=true}
\usepackage{textcomp}
\usepackage[normalem]{ulem}

\DeclareMathOperator{\E}{E}
\DeclareMathOperator{\trace}{tr}

\DeclareMathOperator{\vect}{vec}

\usepackage{setspace}

\newcommand{\nuk}{\newcommand}
\nuk{\EBN}{\begin{eqnarray}{rcl}}
\nuk{\EEN}{\end{eqnarray}}
\nuk{\EBs}{\begin{eqnarray*}{rcl}}
\nuk{\EEs}{\end{eqnarray*}}

\nuk{\EBNp}[1]{\begin{eqnarray}{#1}}
\nuk{\EBsp}[1]{\begin{eqnarray*}{#1}}

\nuk{\ep}{\epsilon}

\nuk{\sg}{\sigma}
\nuk{\lab}{\label}
\nuk{\ben}{\begin{itemize}}

\nuk{\een}{\end{itemize}}
\nuk{\cF}{\mathcal{F}}
\nuk{\cK}{\mathcal{K}}
\nuk{\cO}{\mathcal{O}}

\nuk{\raa}{\rightarrow}
\nuk{\Om}{\Omega}
\nuk{\xih}{\hat{\xi}}
\nuk{\convp}{\xrightarrow{p}}

\newtheorem{thm}{Theorem}

\newtheorem{lem}{Lemma}

\newtheorem{defn}{Definition}

\newtheorem{prop}{Proposition}[thm]
\newtheorem{remark}{Remarks}

\nuk{\Fh}{\hat{F}}
\nuk{\lam}{\lambda}
\nuk{\sgs}{\sigma_\ast}

\newcommand{\EB}{\begin{align*}}
\newcommand{\EE}{\end{align*}}
\nuk{\EQb}{\begin{displaymath}}
\nuk{\ENe}{\end{displaymath}}

\nuk{\lbay}{\left[\begin{array}}
\nuk{\eart}{\end{array}\right]}
\nuk{\abE}[2]{\lbay{c}#1\\#2\\\eart}  %

\nuk{\zit}{\zeta_t}
\nuk{\zitp}{\zeta_{t+1}}
\nuk{\tubytuE}[4]{\lbay{cc}#1 & #2\\ #3 & #4\\\eart}  

\nuk{\ul}{\underline}
\nuk{\Ah}{\hat{A}}
\nuk{\rai}{\rightarrow\infty}

\nuk{\al}{\alpha}
\nuk{\be}{\beta}
\nuk{\del}{\delta}

\nuk{\itum}[1]{\item[(#1)]}
\nuk{\pT}{p_{_T}}
\nuk{\gamT}{\gam_{_{T}}}

\nuk{\Gam}{\Gamma}
\nuk{\pa}{\parallel}
\nuk{\woT}{\frac{1}{T}}

\nuk{\upa}{^\ast}
\nuk{\xiht}{\check{\xi}_t}
\nuk{\xihtp}{\check{\xi}_{t+1}}

\nuk{\Zh}{\check{Z}}
\nuk{\Sh}{\check{S}}

\nuk{\dxioo}{_{\xi,00}}
\nuk{\dxiww}{_{\xi,11}}
\nuk{\dxiwo}{_{\xi,10}}

\nuk{\dxi}{_\xi}
\nuk{\wosqT}{\frac{1}{\sqrt{T}}}
\nuk{\Xh}{\hat{X}}

\nuk{\Qep}{Q_\ep}
\nuk{\Qv}{Q_v}
\nuk{\Qw}{Q_w}

\nuk{\xh}{\hat{x}}
\nuk{\xht}{\xh_t}
\nuk{\Mh}{\hat{M}}

\nuk{\theo}[1]{\begin{theorem}#1\end{theorem}}
\nuk{\lembe}[1]{\begin{lemma}#1\end{lemma}}    
\nuk{\defi}[1]{\begin{definition}#1\end{definition}}
\nuk{\exa}[1]{\begin{example}#1\end{example}}
\nuk{\corbe}[1]{\begin{corollary}#1\end{corollary}}    
\nuk{\propbe}[1]{\begin{proposition}#1\end{proposition}}  
\nuk{\res}[1]{\begin{result}#1\end{result}}
\nuk{\pro}[1]{\begin{proof}#1\end{proof}}
\nuk{\cas}[1]{\begin{cases}#1\end{cases}}
\nuk{\arr}[2]{\begin{array}{#1}#2\end{array}}
\nuk{\rmk}[1]{\begin{remark}#1\end{remark}}
\nuk{\bra}[1]{\left(#1\right)}
\nuk{\sqbra}[1]{\left[#1\right]}
\nuk{\ang}[1]{\langle#1\rangle}
\nuk{\Ver}[1]{\lVert#1\rVert}
\nuk{\ver}[1]{\lvert#1\rvert}
\nuk{\ssum}[1]{{\textstyle\sum}_{#1}}
\nuk{\sprod}[1]{{\textstyle\prod}_{#1}}
\nuk{\inv}[1]{#1^{-1}}

\makeatother

\newcommand{\ESB}[1]{\begin{subequations}\label{#1}\begin{align}}

\makeatletter
\newcommand{\EBn}{\begin{align}}
\newcommand{\EEn}{\end{align}}
\makeatother

\newcommand{\BN}{\begin{enumerate}}
\newcommand{\EN}{\end{enumerate}}
\newcommand{\Bdef}[1]{\begin{defn}\label{def:#1}}
\newcommand{\Edef}{\end{defn}}
\newcommand{\Blem}[1]{\begin{lem}\label{lem:#1}}
\newcommand{\Elem}{\end{lem}}
\newcommand{\Balem}[1]{\begin{alem}\label{lem:#1}}
\newcommand{\Ealem}{\end{alem}}
\newcommand{\Bthm}[1]{\begin{thm}\label{thm:#1}}
\newcommand{\Ethm}{\end{thm}}
\newcommand{\Bprop}[1]{\begin{prop}\label{prop:#1}}
\newcommand{\Eprop}{\end{prop}}
\newcommand{\Bpf}{\begin{pf*}}
\newcommand{\Epf}{{\hfill$\square$}\end{pf*}}

\newcommand{\Soo}{S_{00}}

\newcommand{\Suwo}{S_{u,10}}

\newcommand{\Suoo}{S_{u,00}}
\newcommand{\Suww}{S_{u,11}}
\newcommand{\Bsmat}[1]{\begin{smallmatrix}#1}
\newcommand{\Esmat}{\end{smallmatrix}}
\newcommand{\Bmat}[1]{\begin{matrix}#1}
\newcommand{\Emat}{\end{matrix}}
\newcommand{\Ra}{\Rightarrow}

\newcommand{\xt}{x_t}
\newcommand{\xtpw}{x_{t+1}}
\newcommand{\yt}{y_t}
\newcommand{\ut}{u_t}
\newcommand{\utpw}{u_{t+1}}
\newcommand{\wt}{w_t}

\newcommand{\ept}{\epsilon_t}

\newcommand{\vt}{v_t}

\newcommand{\et}{e_t}

\newcommand{\xit}{\xi_t}
\newcommand{\xitpw}{\xi_{t+1}}
\newcommand{\gam}{\gamma}
\newcommand{\mhaf}{{-\frac12}}
\newcommand{\haf}{{\frac12}}
\newcommand{\lt}{\left}
\newcommand{\rt}{\right}
\newcommand{\Au}{A_u}

\newcommand{\Piu}{\Pi_U}
\newcommand{\Piup}{\Pi_U^\bot}
\newcommand{\Piufp}{\Pi_{\Uf}^\bot}

\newcommand{\uft}{u_{t,f}^+}
\newcommand{\upt}{u_{t,p}^-}
\newcommand{\yft}{y_{t,f}^+}
\newcommand{\ypt}{y_{t,p}^-}
\newcommand{\zpt}{z_{t,p}^-}

\newcommand{\Ab}{\bar A}
\newcommand{\Uf}{U^+_f}
\newcommand{\Yf}{Y^+_f}
\newcommand{\Zp}{Z^-_p}

\newcommand{\Sigpp}{\hat\Sigma_{pp}^\bot}
\newcommand{\Sigfp}{\hat\Sigma_{fp}^\bot}
\newcommand{\Sigff}{\hat\Sigma_{ff}^\bot}

\newcommand{\Sfr}{S$^4$}
\newcommand{\Sfv}{S$^5$}

\begin{document}
\onehalfspacing
\title{Stable State Space SubSpace (S$^5$) Identification}                                                

\author{Xinhui Rong and Victor Solo
\footnote{Authors are with School of Electrical Eng. $\&$ Telecommunications, 
UNSW, Sydney, Australia. }  
}    
  
\maketitle

\begin{abstract}                          
State space subspace algorithms for input-output systems have 
been widely applied but also have a reasonably well-developed
asymptotic theory dealing with consistency.
However, guaranteeing the stability of the estimated system 
matrix is a major issue. 
Existing stability-guaranteed  algorithms are computationally expensive, 
require several tuning parameters, and scale badly to high state dimensions.
Here, we develop a new algorithm that is closed-form and requires no tuning parameters. It is thus computationally cheap and scales easily to high state dimensions.
We also prove its consistency under reasonable conditions.
\end{abstract}

\section*{Copyright Statement}
This work has been submitted to the IEEE for possible publication. Copyright may be transferred without notice, after which this version may no longer be accessible.

\section{Introduction}
\label{intro}
There are two types of linear state space (SS) model 
relevant to system identification.
In input-output SS  (IO-SS) models, the states 
are driven by unmeasured noise and measured input signals.
In state space noise (N-SS) models, 
the state is driven only by an unmeasured noise.

The state space subspace (\Sfr) approach is a powerful
method of SS modelling, applicable to both IO-SS and N-SS models.
It does not rely on parameterization and avoids identifiability issues.
The \Sfr\ method uses projection theorems to 
extract the state or a system-related matrix, 
namely the state sequence or the extended observability matrix, 
from the input and output signals (or only the output signal in the N-SS case) \cite{VanO96}, 
followed by various methods to identify the system matrices. 
The \Sfr\ method is of great interest due to its simple implementation 
and calculation; it involves little more than 
singular value decomposition (SVD). 

\Sfr\ identification with guaranteed system stability  
has long been of interest 
since unstable identification can occur on systems that are known to be stable, 
due to limited measurement time,  or noise. 
The simpler system structure of the N-SS models
allows a wider range of stability-guaranteed algorithms 
as the state equation is a first-order vector autoregression (VAR(1)). 

For N-SS models,
Mari et al. \cite{Mari00} consider a semidefinite programming (SDP) problem 
with Lyapunov stability constraint to 
minimize the weighted discrepancy between 
the unstable initial estimate and the stable one. 
We shall refer to such method the {\it perturbation minimization}. 
{Boots et al. \cite{Boot07} use a constraint generation approach 
which minimizes the norm of the one-step-ahead prediction error under a singular value constraint that is iteratively restricted until a stable solution is reached.}
Tanaka and Katayama \cite{Tana05} also use the perturbation minimization procedure 
{and stabilize the state transition matrix by solving Riccati equations}. 
Jongeneel et al. \cite{Jong23} propose a new LQR problem and 
offer error analysis and statistical guarantees for the VAR(1) problem. 
However, in the N-SS setting, 
the necessity to invert the mean square error sample matrix in the algorithm 
can lead to operational challenges when it approaches singularity 
as is often the case in the simulations. 
We \cite{Rong23} use a {\it forwards-backwards (FB) optimization}
\cite{Burg75}\cite{Stra77}\cite{Nutt77} on the error residuals 
whose solution is closed-form and guaranteed stable for N-SS. 
We further extended the work to encompass symmetry \cite{Rong24i} and rank constraints \cite{Rong24a}. 
More recently, we \cite{Rong24c} developed for VAR(1) 
a class of {\it correlation stable estimators},
which are stable, closed-form and consistent.

For IO-SS models, 
Chui and Maciejowski \cite{Chui96} use an iterative attenuation (IA) method 
where the estimated state sequence or the extended observability matrix 
is iteratively appended until stability is reached. 
However, this method distorts the estimator and introduces additional bias, 
{since the estimator will have a dominating pole with a user-defined magnitude.} 
Van Gestel et al. \cite{VanG01} develop the regularization method 
to the standard least squares to guarantee stability. 
Lacy and Bernstein \cite{Lacy03} extend the SDP method to include input. 
Mallet et al. \cite{Mall08} use a line search method 
based on gradient sampling. 
Umenberger et al. \cite{Umen18} use the expectation maximization (EM) method. 
However, it is too computationally expensive 
since there is a SDP procedure in every iteration of EM. 

More recently another approach to guaranteed stability has appeared.
The idea is to
find the nearest stable matrix to the initial unstable estimate 
\cite{Orba13}\cite{Gill19}\cite{Nofe21}. 
However, they can only guarantee a local minimum and 
no data information is used except for the initial unstable estimate.

Stability-guaranteed algorithms for IO-SS 
can be adapted to N-SS (see \cite{Rong23} for examples), 
but the reverse is not straightforward. 
For example, the SDP methods in \cite{Mari00} and \cite{Lacy03} 
end up being the same 
even though an input is considered in \cite{Lacy03}. 
However, the LQR method in \cite{Jong23} requires 
that the state and noise covariance matrices 
obey a Lyaponuv equation associated with the state transition matrix 
which only occurs in the noise modeling or VAR(1).

In this paper, we consider IO-SS modelling 
with a VAR(1) input process and 
extend the correlation stable estimators 
\cite{Rong24c} to encompass such system structure. 
We  develop 
the first closed-form, tuning-parameter-free stable \Sfr, which we call \Sfv, 
that can be applied to both IO-SS and N-SS. 
We also show statistical consistency  for \Sfv under reasonable conditions.

The rest of the paper is organized as follows. 
We first introduce the IO-SS model in Section \ref{mod}.  
We review canonical variate analysis (CVA) 
subspace identification of the states in Section \ref{S4}, 
and the correlation stable estimators for VAR(1) in Section \ref{svar}. 
We then develop 
\begin{enumerate}
\item[(i)] in Section \ref{S5}, the \Sfv\ algorithm, 
\item[(ii)] in Section \ref{asym}, statistical consistency
for the \Sfv\ estimator. 
\end{enumerate}
We compare \Sfv\ with various previous methods
in simulations in Section \ref{sim} with both low-dimensional
 and high-dimensional examples. 

We define the following notations.  
$A'$ is the transpose of $A$. 
$\rho(A)$ is the spectral radius of $A$, i.e. the largest modulus of the eigenvalues. 
$\Vert A\Vert=\sqrt{\trace(AA')}$ is the Frobenius norm of $A$. 
$\otimes$ is the Kronecker product. 
$I_k$ is the identity matrix with dimension $k$, 
or $I$ the identity matrix with appropriate dimension. 
$\xrightarrow{p}$ means convergence in probability.
w.p.1 means `with probability $1$'.

\section{State Space Modeling}
\setcounter{equation}{0}
\label{mod}
We consider the linear, time-invariant SS model in the innovations form 
\ESB{eq:sse}
\xtpw &= A\xt + B\ut + K\ept\\
\yt &= C\xt + \ept,
\end{align}\end{subequations}
where $\yt$ is the $d$-vector measured output, 
$\ut$ is the $m$-vector measured input, 
$\xt$ is the $n$-vector latent state. 
The innovation $\ept$ is white, 
i.e. $\E[\ept\epsilon_s']=0, t\neq s$, 
with {zero mean and }covariance $\E[\ept\ept']=\Qep>0$. 
{The matrices $A_{n\times n}, B_{n\times m}, C_{d\times n}, K_{n\times d}$ 
contain the system parameters, 
where $K_{n\times d}$ is the Kalman gain. }
We introduce some assumptions for the SS model.

{\bf Assumption A1.}
For the SS model (\ref{eq:sse}), 
\ben
\item The $\ept$ and $\ut$ sequences are statistically independent.
\item $A$ and $A-KC$ are stable $\equiv \rho(A)<1,\rho(A-KC) <1$.
\item $(A,C)$ is observable, $(A, [K, B])$ is reachable. 
\een

Note that $\rho(A-KC)<1$ means the system is minimum-phase, 
and A1 ensures the state space representation (\ref{eq:sse}) is minimal 
\cite[Theorem 2.3.4]{Hann12}.

Further assumptions on $\et$ are needed for the asymptotic analysis below.

{\bf Assumption A2.} For the innovation sequence $\ept$.
\ben
\item $\E[\ept|\cF_{t-1}^{\ep}]=0, \E[\ept\ept'|\cF^{\ep}_{t-1}]=\E[\ept\ept']>0$ 
under the history 
$\mathcal{F}_t^{\ep}=\sigma\{\ept,\epsilon_{t-1},\dotsm\}$. 
\item $\E [\epsilon_{t,a}^4]<\infty, \quad \E[\epsilon_{t,a}\epsilon_{t,b}\epsilon_{t,c}|\cF_{t-1}^{\ep}]=\omega_{a,b,c},$
where $\epsilon_{t,a}$ is the $a$-th entry in $\ept$ and 
$\omega_{a,b,c}$ is a constant independent from $t$. 
\een

We now introduce assumptions on the input sequence. 

{\bf Assumption A3. Input is VAR(1).}\\
$\ut$ obeys the VAR(1) model,
\EBn\label{eq:u}
\utpw = A_{u}\ut + \vt, 
\end{align}
where $A_{u}$ is the $m\times m$ stable VAR(1) matrix, 
and $\vt$ is an iid zero mean white noise sequence
with covariance $Q_v = \E[\vt\vt']>0$. 
We further assume: (i)  $v_t$ satisfies  A2 (with history $\cF^{v}_t$);
and (ii) $A$ and $A_u$ have no common eigenvalues. 

{{\bf Remarks 1. }}
\ben
\itum{i}
Note that in view of A1, the sequences $\ept,\vt$ are statistically independent.
\itum{ii} It is straightforward to show that
any sequence of the form $\wt=
D_1\ept+D_2\vt$ (for constant matrices $D_1,D_2$)
also obeys A2 (with history $\cF_t^{w})$.
\een

Taking variances in (\ref{eq:u}) and assuming stationarity we find
that the variance matrix $\Pi_u\equiv\E[u_tu_t']$ obeys the Lyapunov equation:
$\Pi_u = A_u\Pi_uA_u'+Q_v>0$. 
Further, introducing the lag 1 covariance $\Pi_{u,10} = \E[u_{t+1}u_t']$,
we find $A_u = \Pi_{u,10}\Pi_u^{-1}$.

We aim to estimate 
the system matrices $(A,B,C,K)$, the innovation covariance $\Qep$
and the VAR matrices $\Au,\Qv$ and
require the estimated $A$ and $\Au$ to be stable.

\section{Review of Canonical Variate Analysis \\State Space Subspace Method}
\setcounter{equation}{0}
\label{S4}
The \Sfr\ methods use projections to
estimate the unobserved states \cite{VanO96}, 
which are later used for system identification. 
In this section, we briefly review the canonical variate analysis (CVA) subspace method{\cite{Lari83}. }

First, introduce the lag-$p$ stacked past input vector 
$\upt = [ u_{t-1}'\dots u_{t-p}']'$ 
and the lag-$f$ future input vector 
$\uft = [u_t'\dots u_{t+f-1}']'$. 
Also define similarly the past and future output vector, $\ypt$ and $\yft$ 
and $\zpt = [(\ypt)', (\upt)']'$ 
and $\Ab = A-KC$. 
Using the SS innovation form (\ref{eq:sse}), 
one can easily obtain
\EBn\label{eq:xt}
\xt = \Ab^px_{t-p} + \cK_p \zpt,
\end{align}
where $\cK_p=[K,B,\Ab[K,B],\Ab^2[K,B],\dotsm,\Ab^{p-1}[K,B]]$ 
is the extended controllability matrix and 
has full row rank by the minimality A1 given $p\geq n$. 
Also,
\begin{subequations}\begin{align}
\yft &= \cO_f\xt + \Phi_f\uft + \eta_{t,f}^+\label{eq:yf}\\
	&= \cO_f\cK_p\zpt + \cO_f \bar A^px_{t-p}+\Phi_f\uft + \eta_{t,f}^+,
\end{align}\end{subequations}
where $\cO_f = [C',A'C',\dotsm,(A^{f-1})'C']'$ is the extended observability matrix, 
$\Phi_f$ is lower block triangular and contains the system matrices, and 
$\eta_{t,f}^+$ is related to the future noise and is independent from the remaining terms. 
Also note that $\cO_f\cK_p$ will be of low rank $n$ for sufficiently large $f,p$. 

The minimum-phase assumption in A1 
enables the approximation of $\xt$ 
as a linear transformation of $\zpt$ as 
\EBn\label{eq:xz}
\xt &\approx \cK_p \zpt, 
\end{align}
so that one can write $\yft = \beta_z \zpt + \beta_u\uft + r_{t,f}^+$ 
where $\beta_u$ is some constant matrix, 
$r_{t,f}^+$ is the residual term 
which is in general not orthogonal to $\zpt$ and $\uft$, 
and 
\EBn\label{eq:beta}
\beta_z = \cO_f(\cK_p+\bar A^p D), 
\end{align} 
where $D$ is a finite term. 
However, the minimum-phase assumption 
ensures the bias between $\beta_z$ and $\cO_f\cK_p$ vanishes as $p\to\infty$ 
(at a certain rate, see Section \ref{asym}). 

Motivated by (\ref{eq:xz}), 
CVA predicts $\xt$ as linear combination of $\zpt$ as 
$
\hat x_t = \hat\cK_p\zpt, 
$
and minimizes the noise in {(\ref{eq:yf})}, regressed on $\uft$ as follows. 

Given the input-output data for $t=0,\dotsm,\bar T$, 
choose the lags $p\geq n, f\geq n$ and 
define $T=\bar T- f-p+1$, 
and the past and future data matrices 
$
\Zp = \left[z_{p,p}^-,\dotsm, z_{p+T,p}^-\right],
\Yf = \left[y_{p,p}^+,\dotsm, y_{p+T,f}^+\right],
\Uf = \left[u_{p,p}^+,\dotsm, u_{p+T,f}^+\right]. 
$
{Under A3, $\Uf(\Uf)'$ has full rank w.p.1.} 
We also define the projection matrices 
\EB
\Piu &= U'(UU')^{-1}U\\
\mbox{and } \Piup &= I - \Piu.
\end{align*}
$Y\Piu$ projects the row space of $Y$ onto the row space of $U$ and 
$Y\Piup$ is the residual. 
$\Piup$ has properties such that $(\Piup)^2=\Piup$ and $U\Piup=0$. 

Let $\hat X = [\hat x_p, \dotsm, \hat x_{p+T}]${, which is to be found}. 
Then, by partial regression and (\ref{eq:yf}), 
we estimate $\cO_f$ as 
\EB
\hat\cO_f = \Yf\Piufp\hat X'(\hat X\Piufp\hat X')^{-1}, 
\end{align*}
and predict $\yft$ as
\EB
\hat Y^+_f &= \hat \cO_f\hat X + {\hat\Phi_f}\Uf
\Ra \hat Y^+_f\Piufp = \hat \cO_f\hat X\Piufp, 
\end{align*}
where {$\hat\Phi_f$ is some estimate of $\Phi_f$} 
but is not needed here since it vanishes in the projection. 
Then the optimization problem becomes 
finding $\hat\cK_p$ such that $\hat X = \hat\cK_p\Zp$, and  
$\Vert W^{-\frac12}(\Yf- \hat Y^+_f)\Piufp)\Vert^2$ is minimized, 
where $W$ is a positive definite weighting matrix. 
The solution is given below. 
Let $W=\Sigff=\frac1T\Yf\Piufp(\Yf)', \Sigfp=\frac1T\Yf\Piufp(\Zp)'$ and  $\Sigpp=\frac1T\Zp\Piufp(\Zp)'$. 
Carry out the SVD
\EB
	&(\Sigff)^{-\frac12}\Sigfp(\Sigpp)^{-\frac12} 
= 	\hat U\hat \Lambda \hat V' = [\hat U_{\hat n}, \hat U_*]\left[\Bmat \hat\Lambda_{\hat n} &0\\0&\hat\Lambda_*\Emat\right]\left[\Bmat\hat V_{\hat n}'\\\hat V_*'\Emat\right], 
\end{align*}
where $G^{\frac12}$ is the unique symmetric matrix square root of a positive definite (pd) $G$ and 
$\hat\Lambda_{\hat n}$ contains the largest $\hat n$ singular values in decreasing order down its diagonal, where $\hat n\leq f$ is the estimated state dimension and $\hat U,\hat V$ are normalized singular vectors. 
Then, 
\EBn
\hat\cK_p &= \hat\Lambda_{\hat n}^{\frac12} \hat V_{\hat n}' (\Sigpp)^{-\frac12}\label{eq:cKm}\\
\Ra \hat X &= \hat\cK_p\Zp.\label{eq:Xhat}
\end{align}
We note here for future use that 
$\hat\cO_f=(\Sigff)^{\frac12}\hat U_{\hat n}\hat\Lambda_{\hat n}^{\frac12}$ 
and $\hat\cO_f\hat\cK_p$ is the rank $n$ approximation of 
\EBn\label{eq:betahat}
\hat\beta_z = \Sigfp(\Sigpp)^{-1}, 
\end{align}
the estimator for $\beta_z$ by partial regression without rank constraint. 
Thus, an alternative expression for $\hat\cK_p$ is \cite{VanO96}
$\hat\cK_p = \hat\cO_f^\dagger\hat\beta_z$, 
where 
\EB
\hat\cO_f^\dagger=(\hat\cO_f'(\Sigff)^{-1}\hat\cO_f)^{-1}\hat\cO_f'(\Sigff)^{-1}=\hat\Lambda_{\hat n}^{-\frac12}\hat U_{\hat n}'(\Sigff)^{-\frac12},
\end{align*} 
is the left pseudo inverse of $\hat\cO_f$. This will simplify the discussion of asymptotics a lot.

After getting the estimated states, 
system matrices $A,B,C,{K}$ can be estimated by least squares (LS). 
Let $\hat X_1 = [\hat x_{p+1},\dotsm,\hat x_{p+T}], 
\hat X_0 = [\hat x_p,\dotsm,\hat x_{p+T-1}]$, 
$Y_0 = [y_p,\dotsm,y_{p+T-1}]$, and 
{$U_0 = [u_p,\dotsm,u_{p+T-1}]$.} 
Then, the LS estimators of $A,B,C$ are 
\EBn
\left[\Bmat\hat A_\ast&\hat B\Emat\right] &= 
\hat X_1\left[\Bmat \hat X_0'&U_0'\Emat\right]\left(\left[\Bmat \hat X_0\\U_0\Emat\right]\left[\Bmat \hat X_0'&U_0'\Emat\right]\right)^{-1}\label{eq:ls}\\
\hat C &= Y_0X_0'(X_0X_0')^{-1}. 
\end{align}
Note that $\hat A_\ast$ is not guaranteed to be stable w.p.1. 
For future use we note that
\EBn
\label{eq:A0hat}
\hat A_\ast = \hat X_1\Pi_{U_0}^\bot\hat X_0'(\hat X_0\Pi_{U_0}^\bot \hat X_0')^{-1}.
\end{align}
{Let $\hat E = Y_0 - \hat C\hat X_0$. 
The LS estimator of $K$ is 
$
\hat K = \hat X_1\hat E'(\hat E\hat E')^{-1}.
$ 
The estimator of $\Qep$ is $\woT\hat E\hat E'$. 
}

\section{Correlation Stable VAR(1) Estimators}
\setcounter{equation}{0}
\label{svar}
Our new closed-from stable estimator (\Sfv) builds on 
closed-form stable estimators for VAR(1) models.
These are developed in our recent work \cite{Rong24c} 
and are called {\it correlation stable} (CS) estimators. 

Let the $n$-vector process $\xit$ be a VAR(1) process 
\EBn\label{eq:xit1}
\xi_{t+1} = F\xit + w_t,
\end{align}
with $F$ stable and $\wt$ a zero mean white noise sequence
with covariance $\Qw>0$. 

Then as indicated below A3, under stationarity, the
state covariance satisfies a Lyapunov equation
$\E[\xit\xit']= \Pi\dxi = F\Pi\dxi F' + \Qw>0$.

Introduce $Z = [\xi_p,\xi_{p+1},\dotsm,\xi_{p+T}]$ 
and $Z_0$ to be $Z$ without the last column and 
$Z_1$ to be $Z$ without the first column. 
Also introduce the sample covariance matrices
\EBn\label{eq:Sij}
S_{\xi,ij} = \frac1TZ_iZ_j',\quad i,j\in\{0,1\}.
\end{align}
The CS estimator is given in the following theorem. 

{\bf Theorem 1. }\cite{Rong24c}
{Correlation Stable (CS) Estimator for VAR(1). } 
{Suppose that $\xit$ satisfies the stable VAR(1) given in (\ref{eq:xit1}) 
and that $\wt$'s are iid with mean $0$, covariance $\Pi\dxi>0$ and finite fourth moment.} 
Let $P$ be any positive definite consistent estimator of $\Pi\dxi$. 
The following estimator of $F$ is stable and consistent. 
\EBn
\hat F_{P} = P^{\frac12}RP^{-\frac12}\convp F,\label{eq:FhatP}
\end{align}
where $R=S\dxiww^{-\frac12}S\dxiwo S\dxioo^{-\frac12}$ is the correlation matrix. 

{\it Proof. }
 \cite[Theorem IVb]{Rong24c} 
For completeness and later extension,
we repeat part of the proof here. 
Firstly note that the eigenvalues
of $\hat F_P$ are the same as those of $R$
since $|\lam I-\hat F_P|=|\lam I-R|$.
Next $\rho(R)\leq \sg(R)\equiv$ largest singular value of $R$ {\cite[Chapter 4]{Horn91}}. 
But by the Cauchy-Schwartz inequality{\cite[Theorem 11.2]{Magn19}}  $\sg(R)<1$ w.p.1.
The {stability} now follows.
\hfill$\square$

{\bf Corollary 1}. Setting $P=S\dxiww$ we get
\EBn
\hat F =S\dxiwo S\dxioo^{-\frac12} S\dxiww^{-\frac12}\convp F.\label{eq:Fhat}
\end{align}

\section{Stable State Space Subspace Identification}
\setcounter{equation}{0}
\label{S5}
We now consider extending the CS estimator 
to IO-SS models.

\subsection{Markovian State Extraction}
First note that the VAR(1) in A3 can be written as a SS model as follows
\EQb
\zitp=\Au\zit+\Au{v_{t-1}}\mbox{ and }\ut=\zit+{v_{t-1}}
\ENe
where $\zit$ is an unobserved state.

We can now form a 
 joint SS model as follows. 
\ESB{eq:jSS}
\lt[\Bmat\xtpw\\\zitp\Emat\rt] 
&= 
\lt[\Bmat A&B\\0&A_u\Emat\rt]\lt[\Bmat\xt\\\zit\Emat\rt]+\lt[\Bmat K\ept\\ \Au {v_{t-1}}\Emat\rt]\\
\abE{\yt}{\ut} &= \lt[\Bmat{C}&{0}\\{0}&{I}\Emat\rt]\lt[\Bmat{\xt}\\{\zit}\Emat\rt]+ \lt[\Bmat{\ept}\\{v_{t-1}}\Emat\rt].
\end{align}\end{subequations}

Since we now have a noise driven SS model, we could try to apply
the stability guaranteed method we developed in \cite{Rong24c}.
But that method cannot handle the constraint that the left
bottom corner block entry in the system matrix is $0$.
We have instead been led to find a different approach.

It turns out to be possible to construct a new state whose
transition matrix is $A$.

{\bf Result I}.
Assume $\xt,\ut$ satisfy (\ref{eq:sse}) and (\ref{eq:u}) 
and the assumptions A1-A3 are fulfilled. 
Introduce $M_{n\times m}$ which solves the Sylvester equation 
\EBn\label{eq:M}
AM - MA_u + B = 0. 
\end{align}
Then $\xit=\xt - Mu_t$ satisfies a VAR(1) taking the form 
\EBn\label{eq:xit}
\xitpw = A\xit + w_t,\quad w_t = K\ept-M v_t.
\end{align}

{\bf Remarks 2.} 
\BN
\item[(i)] From now on we 
will be using the notation and results of Section IV so 
are effectively setting $F=A$.
\item[(ii)]
The Sylvester equation (\ref{eq:M}) has a unique solution $M$ 
iff $A$ and $A_u$ do not have common eigenvalues \cite[Ch.8]{Datt04}.
This is guaranteed by A3.
\item[(iii)] $w_t$ is white and not correlated to $\xi_t$, so $\xi_t$ is bona-fide 
Markovian state sequence.
Note that $\wt$ has covariance $\Qw=K\Qep K' + M\Qv M'\geq0$.
\item[(iv)] In steady state, the covariance $\Pi\dxi=\E[\xi_t\xi_t']$ 
satisfies a Lyapunov equation 
$
\Pi\dxi = A\Pi\dxi A' + \Qw.
$
\EN
{\it Proof. } 
We have 
\EB
\xitpw - A\xit 
= &A\xt + B\ut+K\ept - Mu_{t+1} - A\xt + AMu_t\\
= &B\ut  + K\ept - M(A_uu_t + v_t)+ AM u_t\\
= &(B - MA_u + AM)u_t + K\ept - Mv_t\\
= &K\ept - Mv_t=w_t,
\end{align*}
as quoted.\hfill$\square$

The new algorithm is enabled by the following result.

{\bf Result II}.
$(A,\Qw^{\frac12})$ is reachable.

{\it Proof. }
Suppose $(A,\Qw^{\frac12})$ is not reachable. 
Then there exists left eigenvector $v$ of $A$
with eigenvalue  $\lam$, i.e. $v'A=\lam v'$ so that 
$v'\Qw^{\frac12}=0$. 
But then $v'\Qw v=v'K\Qep K'v + v'M\Qv M'v=0$.
But $\Qep,\Qv$ are positive definite so we must have $v'K=0,v'M=0$.
 Left multiplying by $v'$ in the Sylvester equation (\ref{eq:M}) yields 
$v'AM-v'MA_u+v'B=0\Ra v'B=-\lambda v'M=0$.
So $v'B=0,v'K=0$ 
which contradicts the reachability of $(A,[K,B])$ in A1. The result follows.
\hfill$\square$

{\bf Remark 3.} In view of Remark 2(iv),
by properties of the Lyapunov equation 
\cite[Lemma D.1.2(v)]{Kail00} and Result II we
conclude that $\Pi_\xi$ is positive definite.

\subsection{The Closed-Form Stable Estimators}
We now develop the closed-form stable estimators for both $A_u$ and $A$. 
We first extract an estimator $\xiht$ of the latent Markovian state 
and then apply the correlation stable estimators as follows.

{\bf \Sfv Algorithm}.\\
Closed-form  Stable State Space SubSpace Algorithm. \\
The closed-form algorithm is in two parts.

(I) $\Ah_u$: The closed-form stable estimator for $A_u$: \\ 
Let {$\bar U = [u_0,\dotsm,u_{\bar T}]$ and $\bar U_1$ be $\bar U$ without the first column 
and $\bar U_0$ be $\bar U$ without the last column.} Define $S_{u,ij} = \frac1{\bar T}\bar U_i\bar U_j',i,j\in\{0,1\}$. 
Then
\EBn\label{eq:Auhat}
\hat A_{u} =S_{u,10}S_{u,00}^{-\frac12}S_{u,11}^{-\frac12},
\end{align}
is a CS estimator of $\Au$.

(II) $\Ah$: The closed-form stable estimator for $A$: \\ 
Let $\hat X = [\hat x_p,\dotsm,\hat x_{p+T}]\in\mathbb{R}^{\hat n\times (T+1)}$ 
be the state estimators from \Sfr\, 
and $\hat A_\ast, \hat B$ be the LS estimators given in (\ref{eq:ls}), 
where $\hat A_\ast$ may be unstable. 
Then solve for $\hat M$ the Sylvester equation 
\EBn\label{eq:Mhat}
\hat A_\ast \hat M - \hat M\hat A_u + \hat B=0.
\end{align}
Now construct the estimated transformed state sequence
\EBn\label{eq:xihat}
\xiht = \hat x_t - \hat Mu_t, 
\end{align} 
for $t = p,\dotsm,p+T$. 
Define $\Zh,\Zh_1,\Zh_0$ and $\Sh_{\xi,ij}$ 
analogously to $Z,Z_1,Z_0$ and $S_{\xi,ij}$ in (\ref{eq:Sij}) , respectively, 
but replacing $\xit$ by $\xiht$. 
Then the CS estimator of $A$ is given by
\EBn\label{eq:Ahat}
\hat A=\Sh\dxiwo\Sh\dxioo^{-\frac12}\Sh\dxiww^{-\frac12},
\end{align}

{\bf Remarks 4. }
\BN
\item[(i)] 
The stability of $\hat A_u$ 
follows directly from Theorem 1 
since $u_t$ is observed. 
The stability of $\hat A$ also follows since replacing the true states with the estimated ones 
does not affect the proof of stability in Theorem 1. %
\item[(ii)] 
The computational cost of solving
the Sylvester equation is minor.
\item[(iii)] 
The Sylvester equation (\ref{eq:Mhat}) has a unique solution iff $\hat A_\ast$ and 
$\hat A_{u}$ have no common eigenvalues. 
This holds w.p.1 because, as proved below,
they are consistent estimators so they can 
not accumulate mass other than at the true eigenvalues which are distinct by A3.
\EN


\section{Asymptotic Analysis}
\setcounter{equation}{0}
\label{asym}
In this section, we develop the {strong} consistency 
for  $\hat A_u$ and $\hat A$. 
Since $\hat B,\hat C,\hat K$ are least squares
estimators, their consistency will follow
from the works we cite below, i.e. \cite{Baue05}, \cite{Pete96}. 
We now obtain, for the first time, {strong} consistency results for
a guaranteed stable system matrix estimator of an IO-SS system.

Throughout this section, 
we assume 
\ben
\item[(a)] the true model order $n$ is known{, i.e. $\hat n=n$}.
\item[(b)] the past lag $p\to\infty$ at a certain rate (see below). 
\een

We introduce the big-O notation. 
A function $g_{_T}=O(\gamma_{_T})$ if there exists a constant $c$  
such that $\limsup_T |g_{_T}|/\gamma_T\leq c<\infty$ holds w.p.1. 
We also introduce 
$\gamT= \sqrt{\log\log T/T}.$

\subsection{Preliminary Consistency Results} 
Our theorems rely on two fundamental sets of results for LS estimators as follows.

{\bf Theorem 2}.\cite{Hann12}
\label{lem:conu}
Consider the VAR(1) input signal obeying A3.
Introduce the LS estimator of $A_u$, namely, 
$\hat A_{u*} = \Suwo\Suoo^{-1}$.
Then the following results hold w.p.1. 
\BN
\item[(a)] Any eigenvalue $\lambda_0$ of $\Suoo$ satisfies \\
$0<a_0<\lambda_0<b_0<\infty$, for some constants $a_0,b_0$. 
\item[(b)] Any eigenvalue $\lambda_1$ of $\Suww$ satisfies \\
$0<a_1<\lambda_1<b_1<\infty$, for some constants $a_1,b_1$.  
\item[(c)] $\Vert\Suoo\Vert=O(1), \Vert\Suww\Vert = O(1), \Vert\Suwo\Vert=O(1)$.
\item[(d)] $\Vert\hat A_{u*}\Vert = O(1)$. 
\EN
Further, the following strong consistency results hold w.p.1. as $\bar T\to\infty$. 
\BN
\item[(e)] $\Vert \Suoo - \Pi_u\Vert = O(\gam_{\bar T}), 
\Vert \Suww - \Pi_u\Vert = O(\gam_{\bar T})$, \\
$\Vert \Suwo - \Pi_{u,10}\Vert = O(\gam_{\bar T})$.
\item[(f)] $\Vert \hat A_{u*} - A_u\Vert\leq O(\gam_{\bar T})$. 
\EN

{\bf Corollary 2}. 
Results (a),(b),(c),(e) also hold for $\xit$ of Theorem 2,
i.e. $S_{\xi,00},S_{\xi,11},S_{\xi,10}$ obey the corresponding results.

{\it Proof}. This follows from Remarks 1(ii) following
assumption A3 from which it follows that $\xit$ obeys A2.

{\bf Theorem 3}. \cite{Pete96}
\label{lem:conx}
Suppose the input process $\ut$ is generated by (\ref{eq:u}) satisfying assumption A3 
and the output process $\yt$ is generated by (\ref{eq:sse}) satisfying assumption A1. 
Also assume that the true model order is known such that $\hat n =n$ 
and in the \Sfr\ steps, the future lag $f\geq n$ is fixed and finite, and 
the past lag $p=\pT$ is chosen such that 
for some finite $\alpha$,
$\pT/(\log T)^\alpha\to0$ w.p.1, as $\pT\rai$.
Introduce $\Gamma = \hat\cO_f^\dagger\cO_f$. 
The following results hold w.p.1. 
\BN
\item[(a)] $\Vert \hat A_*\Vert=O(1), \Vert \hat B\Vert = O(1)$, \\
$\Vert\Gamma\Vert=O(1), \Vert\Gamma^{-1}\Vert=O(1)$. 
\EN
Further, 
the following strong consistency results hold w.p.1. as $T\to\infty$. 
\BN
\item[(b)] $T^\mhaf\Vert\hat X-\Gamma X\Vert\leq \pT O(\gamT)$.
\item[(c)] $\Vert\hat A_*-\Gamma A \Gamma^{-1}\Vert\leq \pT O(\gamT)$.
\item[(d)] $\Vert \hat B - \Gamma B\Vert\leq \pT O(\gamT)$. 
\EN

{\bf Remark 5. } In \cite{Pete96}, the results are developed 
when $f_{T}=\pT\to\infty$ under the rate specified above. 
Results for finite $f$ can be proved without difficulty. 

\subsection{$\hat A_u$ Consistency} 
We now develop consistency for the closed-form stable estimator $\hat A_u$.

{\bf Result III}. Strong Consistency of $\hat A_u$.
\label{thm:conAu}
Under A3,
$\Vert\hat A_u - A_u\Vert\leq O(\gam_{\bar T})$ 
w.p.1 as $\bar T\to\infty$.

{\it Proof. }
We have $\Vert \hat A_{u}-A_u\Vert\leq\Vert \hat A_{u}-\hat A_{u*}\Vert + \Vert \hat A_{u*}-A_u\Vert$. 
But by Theorem 2(f), $\Vert \hat A_{u*}-A_u\Vert\leq O(\gam_{\bar T})$. 
We now show  that $\Vert \hat A_{u}-\hat A_{u*}\Vert \leq  O(\gam_{\bar T})$ yielding the result. 
We have
\EB 
\Vert \hat A_{u}-\hat A_{u*}\Vert &= \Vert \hat A_{u*} S_{u,00}^{\frac12}\Suww^{-\frac12} - \hat A_{u*}\Vert\\
	&= \Vert\hat A_{u*}(\Suoo^\haf-\Suww^\haf)\Suww^\mhaf\Vert\\
	&\leq\Vert \hat A_{u*} \Vert\Vert S_{u,00}^{\frac12}-\Suww^{\frac12}\Vert\Vert \Suww^{-\frac12}\Vert 
\end{align*}
Applying a perturbation bound on matrix square roots \cite[Lemma 2.2]{Schm92} we find
\EB
\Vert S_{u,00}^{\frac12} - \Suww^{\frac12}\Vert &\leq \frac1{a_0+a_1} \Vert S_{u,00}-\Suww\Vert\\
	&\leq  \frac1{a_0+a_1} (\Vert\Suoo-\Pi_u\Vert + \Vert\Suww-\Pi_u\Vert)\\
	&=O(\gam_{\bar T}), 
\end{align*}
where $a_0>0$ and $a_1>0$ uniformly bound the eigenvalues of 
$S_{u,00}$ and $\Suww$ from below, respectively, in view of 
Theorem 2(a,b). 
Note that by Theorem 2(c,d), $\Vert \hat A_{u*}\Vert=O(1)$, 
$\Vert\Suww\Vert=O(1)\Ra\Vert\Suww^{-\frac12}\Vert=O(1)$. 
It follows that $\Vert \hat A_{u}-\hat A_{u*}\Vert\leq O(\gam_{\bar T})$. 
\hfill$\square$

\subsection{$\hat A$ Consistency} 
To proceed we need some further preliminary results. 

{\bf Result IV}. \label{lem:conZ} 
Under A1-A3,
the following strong consistency results hold w.p.1. as $T\to\infty$. 
\BN
\item[(a)] $\Vert\hat M-\Gamma M\Vert \leq \pT O(\gam_{_T})$.
\item[(b)] $T^\mhaf\Vert\Zh-\Gamma Z\Vert\leq \pT O(\gam_{_T})$. 
\item[(c)] $\Vert \Sh\dxioo-\Gamma\Pi\dxi\Gamma'\Vert\leq \pT O(\gam_{_T})$\\
$\Vert \Sh\dxiww-\Gamma\Pi\dxi\Gamma'\Vert\leq \pT O(\gam_{_T})$, \\
$\Vert \Sh\dxiwo-\Gamma\Pi_{\xi,10}\Gamma'\Vert\leq \pT O(\gam_{_T})$.
\EN

{\it Proof. }

\ul{(a)}
Let $\hat G = I_m\otimes\hat A_* -\hat A_u\otimes I_n$ 
and $G = I_m\otimes \Gamma A\Gamma^{-1} - A_u\otimes I_n$. 
Using the vec operator solution to the Sylvester equation, we find
\EB
\Vert\hat M - \Gamma M\Vert &= \Vert \vect(\hat M-\Gamma M)\Vert \\
	&= \Vert\hat G^{-1}\vect(\hat B) - G^{-1}\vect(\Gamma B)\Vert\\
	&\leq\Vert\hat G^{-1} - G^{-1}\Vert\Vert\hat B\Vert + \Vert G^{-1}\Vert\Vert\hat B-\Gamma B\Vert.
\end{align*}
However, 
$\Vert\hat B\Vert =O(1),\Vert G^{-1}\Vert = O(1), 
\Vert\hat B-\Gamma B\Vert \leq \pT O(\gam_{_T})$ 
by Theorem 3(a,d) and Theorem 2(d). 
Next
\EB
\Vert\hat G^{-1}-G^{-1}\Vert &= \Vert \hat G^{-1}(\hat G - G)G^{-1}\Vert\\
	&\leq \Vert\hat G^{-1}\Vert\Vert\hat G-G\Vert\Vert G^{-1}\Vert, 
\end{align*}
where $\Vert G^{-1}\Vert = O(1), \Vert \hat G^{-1}\Vert=O(1)$. Now we write
\EB
\Vert \hat G-G\Vert &= \Vert I_m\otimes
(\hat A_*-\Gamma A\Gamma^{-1}) - (\hat A_u-A_u)\otimes I_n\Vert\\
	&\leq \Vert I_m\otimes(\hat A_*-\Gamma A\Gamma^{-1})\Vert 
	+ \Vert(\hat A_u-A_u)\otimes I_n\Vert\\
	&= \sqrt m\Vert \hat A_*-\Gamma A\Gamma^{-1}\Vert 
	+ \sqrt n\Vert\hat A_u-A_u\Vert\\
	&\leq \pT O(\gam_{_T}).
\end{align*}
Combining these bounds yields result (a). 

\ul{(b)} 
We have $\xiht=\xht-\Mh\ut$ and $\xit=\xt-M\ut$. 
Thus, 
$\Zh=\Xh-\Mh U$ and $Z=X-MU${, where $U=[u_p,\dotsm,u_{p+T}]$}. So
\EBs
\Zh-\Gam Z&=&\Xh-\Gam X-(\Mh-\Gam M)U\\
\Ra
\wosqT\Vert\Zh-\Gam Z\Vert&\leq&\wosqT\Vert\Xh-\Gam X\Vert+\Vert\Mh-\Gam M\Vert\frac{\Vert U\Vert}{\sqrt{T}}\\
&\leq&
\pT O(\gamT)+\pT O(\gamT)\frac{\Vert U\Vert}{\sqrt{T}}
\EEs
by Theorem 3(b) and result (a). Continuing, via Theorem 2
\EBs
\frac{\Vert U\Vert^2}{T}&\leq&\woT\Vert S_{u,11}-\Pi_u\Vert
+\Vert\Pi_u\Vert+\woT\Vert u_0\Vert^2\\
&\leq& \pT O(\gamT)+O(1)+O(1/T)=O(1), 
\EEs
and the quoted bound follows.

\ul{(c)} Consider $\Sh\dxioo$ for example. 
First write
\EBn\label{eq:soo}
&\Vert\Sh\dxioo - \Gamma\Pi_\xi\Gamma'\Vert\nonumber \\
 =&\Vert(\Sh\dxioo - \Gamma S\dxioo\Gamma')
+(\Gamma S\dxioo\Gamma' - \Gamma\Pi_\xi\Gamma')\Vert\nonumber\\
	\leq &\Vert \Sh\dxioo - \Gamma S\dxioo\Gamma'\Vert + \Vert\Gamma\Vert^2 \Vert S\dxioo-\Pi_\xi\Vert.
\end{align}
We need to show both norms decay to $0$. 
For the first norm, write
\EB
	\Sh\dxioo-\Gamma S\dxioo\Gamma'
=	&\frac1T(\hat  Z_0\hat  Z_0' -\Gamma  Z_0 Z_0'\Gamma')\\
=	&\frac1T(\hat Z_0-\Gamma Z_0)(\hat Z_0-\Gamma Z_0)' \\
	&+\frac1T(\Gamma Z_0\hat  Z_0' + \hat  Z_0 Z_0'\Gamma' - 2\Gamma Z_0 Z_0'\Gamma').
\end{align*}
We have $\Vert\frac1T(\hat Z_0-\Gamma Z_0)(\hat Z_0-\Gamma Z_0)'\Vert \leq \pT^2O(\gam_{_T}^2)$ 
by result (b). For the remaining terms, we have
\EB
\Vert\frac1T(\hat Z_0 Z_0'-\Gamma Z_0 Z_0')\Gamma'\Vert 
= 	&\Vert T^{-\frac12}(\hat  Z_0
-\Gamma Z_0)T^{-\frac12} Z_0'\Gamma'\Vert\\
\leq	&\Vert T^{-\frac12}(\hat  Z_0-\Gamma Z_0)\Vert\Vert T^{-\frac12}Z_0\Vert\Vert\Gamma\Vert. 
\end{align*}
We have 
$\Vert T^{-\frac12}(\hat  Z_0-\Gamma Z_0)\Vert\leq \pT O(\gam_{_T})$
 by result (b) and 
$\Vert\Gamma\Vert=O(1)$ by Theorem 3(a). 
But $\Vert T^\mhaf Z_0\Vert=\Vert T^\mhaf
 (\hat X_0 - \hat MU_0)\Vert\leq\Vert T^\mhaf \hat X_0\Vert 
 + \Vert \hat M \Vert \Vert T^\mhaf U_0\Vert$. 
We have $\Vert\hat M\Vert = O(1)$, 
because norms of $\hat A_*,\hat A_u, \hat B$ are all bounded, 
and also $\Vert T^\mhaf U_0\Vert=O(1)$ by Theorem 2(c). 
Now note that $\hat X = \hat\cK_p Z_p^-$. 
In \cite{Pete96}, it is shown that $\Vert\hat\cK_p\Vert=O(1)$ (above their eq.39), 
and $T^{-1}Z_p^-(Z_p^-)'=O(1)$ (their eq.28). 
Thus, $\Vert T^\mhaf \hat X_0\Vert=O(1)$ 
so the above norm decays at a rate of $\pT O(\gam_{_T})$. 
It then follows that 
$\Vert\Sh\dxioo - \Gamma S\dxioo\Gamma'\Vert \leq \pT O(\gam_{_T})$. 

Now for the second norm in (\ref{eq:soo}), note 
that $\xi_t$ is a VARMA process generated by the 
joint innovation $[\ept',v_t']'$ which satisfies A2. 
Then $\Vert\frac1{T}\ssum{t=p}^{p+T-1} \xi_{t+i} \xi_{t}' - \E[ \xi_{t+i} \xi_t']\Vert\leq O(\gam_{_T})$ w.p.1, 
for $|i|\leq(\log T)^\alpha$, for any $\alpha$ 
(see \cite{Hann12} or \cite[eq.26]{Pete96}). 
Then taking $i=0$, we have $\Vert S\dxioo-\Pi_\xi\Vert\leq O(\gam_{_T})$. 
Combining the above yields the result for $\Sh\dxioo$. 
Results for $\Sh\dxiwo$ and $\Sh\dxiww$ follow analogously. 
\hfill$\square$

We can now study the consistency of $\hat A$.

{\bf Result V}. Consistency of $\hat A$.\\
\label{thm:conA}
Assume A1,A2,A3 hold.
Also assume:
\BN
\item[(i)]
the true model order is known so that $\hat n =n$;
\item[(ii)]
the future lag $f\geq n$ is fixed and finite;
\item[(iii)] 
the past lag $p=p_{_T}$ is chosen such that $p_{_T}/(\log T)^\alpha\to0$ w.p.1, 
as $p_{_T}\rai$,
for some finite $\alpha$. 
\EN
Then, $\Ah\to A$ w.p.1. Further, 
$\Vert \hat A - \Gamma A\Gamma^{-1}\Vert\leq \pT O(\gam_{_T})$, 
w.p.1, as $T\to\infty$. 

{\it Proof. } 
Decompose $\hat A - \Gamma A \Gamma^{-1} =\al+\be-\del$, where
\EB
\al &= (\Sh\dxiwo-\Gamma\Pi\dxiwo\Gamma')(\Gamma\Pi_\xi\Gamma')^{-1}\\
\be&= \Sh\dxiwo[\Sh\dxioo^{-1}-(\Gamma\Pi_\xi\Gamma')^{-1}]\\
\del&= \Sh\dxioo^{-\frac12}(\Sh\dxioo^\mhaf-\Sh\dxiww^\mhaf).
\end{align*}
Then use the norm inequalities 
and Result IV(c)
and take analogous steps to treat the matrix inverse and square root 
as in previous proofs to get the quoted result. \hfill$\square$

{\bf Remarks 6. }
\begin{enumerate}
\item[(a)]  The condition that the future lag $f$ also tends to infinity is a stronger assumption, 
under which the same results hold. 
\item[(b)] The stable, closed-form estimator in \cite{Maci95} is shown 
to be consistent when both $f\to\infty, p\to\infty$ but it has an asymptotic bias when $f$ is finite
{\cite[Theorem 3]{Baue05}}. 
\end{enumerate}

\section{Simulations}
\label{sim}

We compare both the precision and computational effort of our S$^5$ 
against the existing algorithms that guarantee system stability, 
with a SS model order $n$ up to $1024$. 
We omit discussions about $\hat A_u$ 
and concentrate here on the IO-SS estimates 
since $\hat A_u$ is a pure VAR(1) stable estimator  
and has already been discussed in simulations in \cite{Rong24c}.

In \cite{Rong23}, we compared our 
FB estimator for N-SS models 
against various algorithms and found 
the SDP \cite{Mari00} to be the most competitive. 
Thus, we include here the SDP method, 
but we use a version for IO-SS \cite{Lacy03}.
More recent developments include 
the LQR method \cite{Jong23} 
and the category of `nearest-stable' algorithms: 
Orth\cite{Nofe21}, FGM\cite{Gill19}, SuccConv\cite{Orba13}. 
However, the LQR method only treats 
a VAR(1) process without an input 
and needs the noise covariance to be pd. 
Thus, we are unable to apply 
this method to our problem. 
The `nearest-stable' algorithms are developed 
without system context 
and are not for SS models. 
We also found some fundamental 
problems for such methods (see below). 
However, because they are the 
{more recent} stability-guaranteed algorithms 
and also because it is not hard 
to amend them to fit a SS model, 
we feel compelled to include them in the simulations. 

Thus, we will compare our (a) \Sfv\ to 
(b) Orth, (c) FGM, (d) SDP. 

We emphasize here that expect our \Sfv\ which is closed-form, 
all algorithms have multiple tuning parameters. \\
$\bullet$ Orth uses Riemannian trust-regions solver 
whose tuning parameters include initial iterate, convergence tolerance, trust-region radiuses, 
accept/reject threshold, etc. \\
$\bullet$ FGM has tuning parameters: initial iterate (very sensitive), 
eigenvalue bounds, fast gradient method step size, line-search parameters, etc. \\
$\bullet$ SDP has tuning parameters: left and right weighting matrices, 
scaling parameter in the Lyapunov equation, etc.

Besides, more needs to be addressed 
about the `nearest-stable' algorithms.

\subsection{Pre-Simulation Discussions on the `Nearest-Stable' Algorithms}
The `nearest-stable' algorithms seek to minimize $\Vert A_s-\hat A_*\Vert$, 
the Frobenius distance between a given unstable matrix $\hat A_*$ 
to the stable $A_s$ in the (non-convex {and non-smooth}) set of stable real matrices. 
Some problems have already been pointed out 
by their authors, e.g. they are only able to find a local minimum
\cite{Nofe21}\cite{Gill19}\cite{Orba13} 
and some algorithms scale badly to higher matrix dimensions \cite{Nofe21}. 
However, what lacks discussion is their ignorance of the system context. 
We address three main points.

{\it (1) Incorrect weighting: } 
Consider the one-step mean square error 
$\Vert Z_1-A_sZ_0\Vert = \Vert Z_1-\hat A_*Z_0+(\hat A_*-A_s)Z_0\Vert
\leq\Vert Z_1-\hat A_*Z_0\Vert+\Vert(\hat A_*-A_s)Z_0\Vert$. 
Then the perturbation should be weighted as $\Vert(\hat A_*-A_s)\Soo^{\frac12}\Vert$.

{\it (2) Distorted poles: }
As discussed in \cite{Nofe21}, 
the `nearest-stable' algorithms tend to put 
the dominating pole very close to the unit circle, e.g. at $1-10^{-10}$), 
distorting pole locations and risking instability due to rounding errors.
Solving a Lyapunov equation $P=A_sPA_s'+Q$ also needs extra attention in this case. 

{\it (3) Orthogonal similarities: }
Different `nearest-stable' algorithms yield solutions 
with nearly identical perturbation norms 
but vastly different orientations. 
This inconsistency can affect the basis of the estimated 
$A$ in SS identification. 
For example, the unstable matrix $\lt[\Bmat2&2&2\\2&2&2\\2&2&2\Emat\rt]$, 
considered in \cite{Nofe21} and \cite{Gill19}, 
has solutions: 
\begin{itemize}
\item \cite{Nofe21} a seemingly global minimizer 
$\lt[\Bmat1 &2 &2\\
0 &1 &2\\
0 &0 &1\Emat\rt]$\\
with objective norm $15$ , 
\item \cite{Nofe21} the Orth solution $\lt[\Bmat1.0000 &0.9550 &1.5848\\
1.0450 &1.0000 &2.6297\\
0.4152 & -0.6297 &1.0000\Emat\rt]$\\
with objective norm $15+10^{-15}$ , and 
\item \cite{Gill19} the FGM solution $\lt[\Bmat0.9969 &1.4010 &0.7688\\
0.5544 &0.9878 &-0.6507\\
1.2476& 2.6740 &1.0112\Emat\rt]$ \\
with objective norm $15.02$ .
\end{itemize}
Despite minimal differences in perturbation norms, 
these matrices are oriented differently, 
causing inconsistencies in SS identification estimates.

These weaknesses of the `nearest-stable' algorithms 
will be exposed in the following simulations.

\begin{figure*}[t]
\begin{center}
\includegraphics[width=15cm]{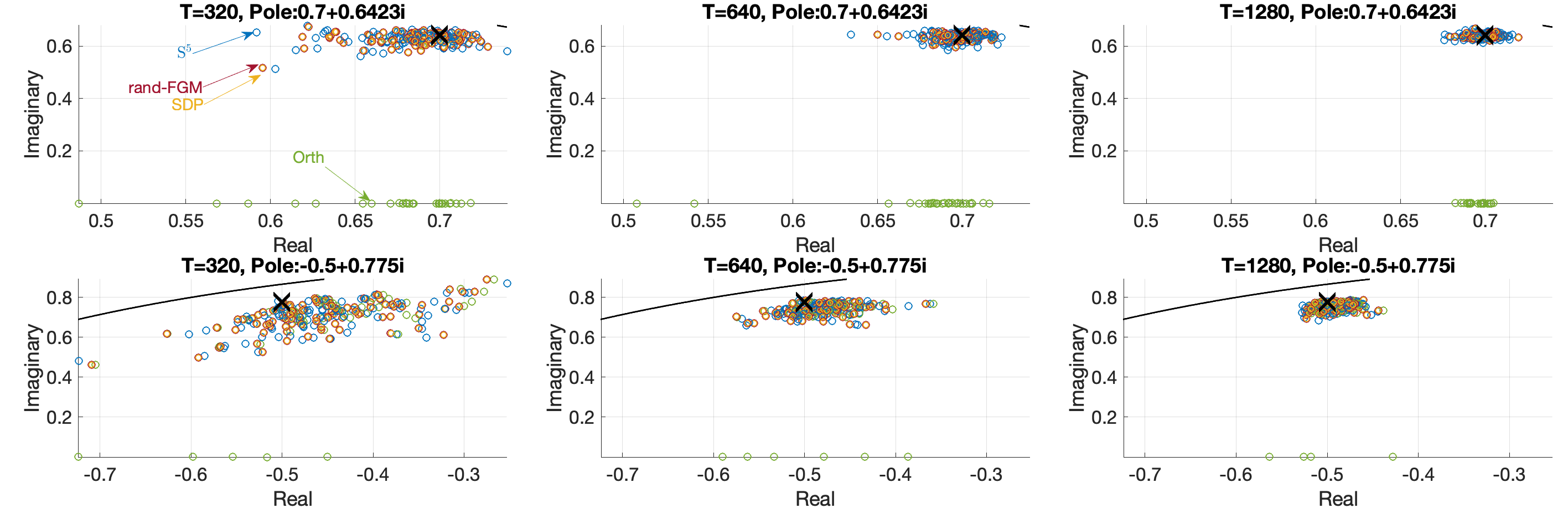}
\caption{Locations of the estimated complex poles  for $n=5$: 
Poles for \Sfv,  rand-FGM and SDP cluster closely 
around the true ones. Poles for rand-FGM and SDP almost overlap. 
However, a large number of poles for Orth lie on the real axis 
yielding a large discrepancy.}
\label{fig:poleloc}
\end{center}
\end{figure*}

\begin{figure}[t]
\begin{center}
\includegraphics[width=12cm]{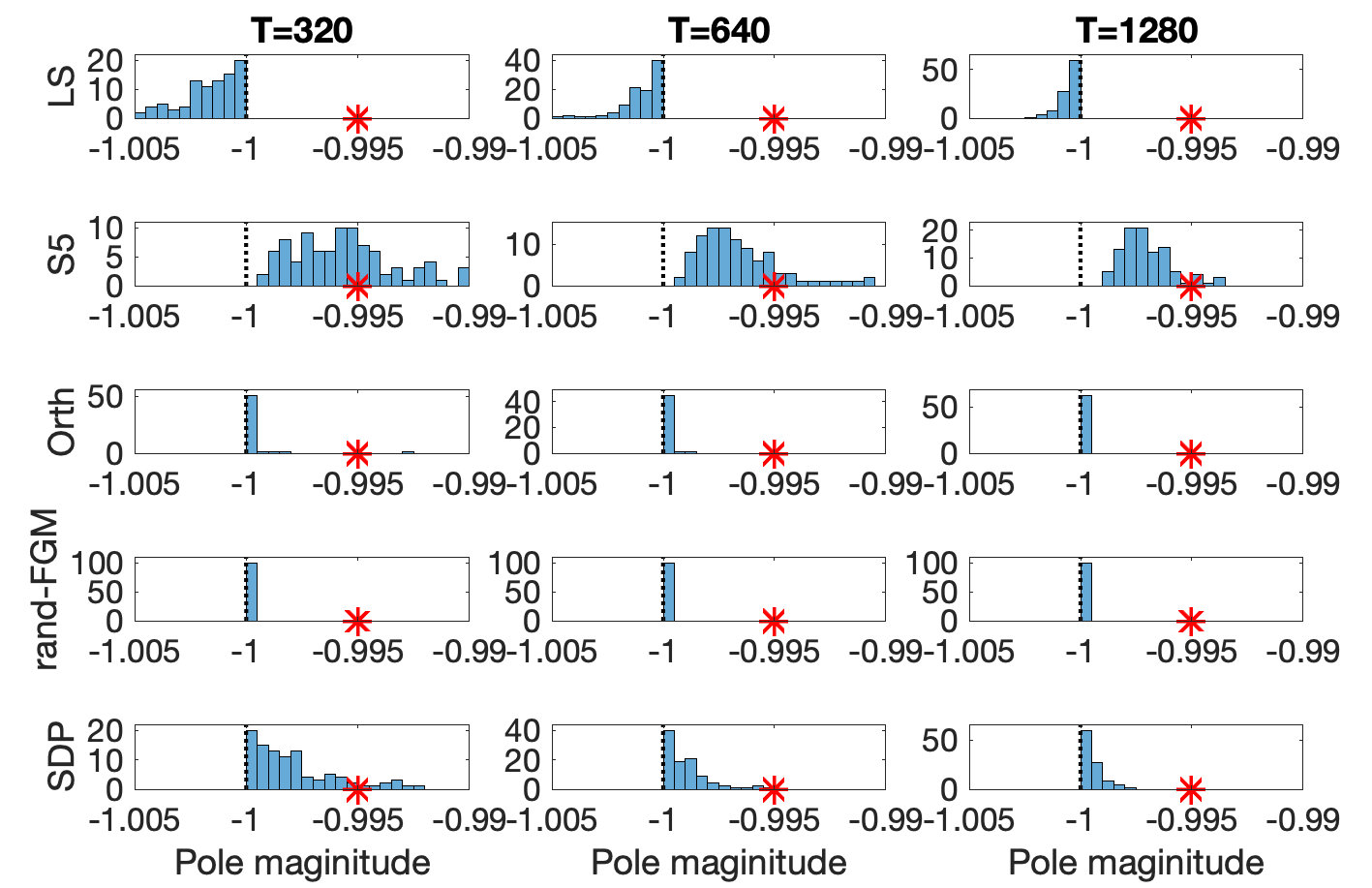}
\caption{Histograms of the estimated real dominating poles for $n=5$: 
We select LS estimates that are unstable. 
The `$*$' marks are the true dominating pole. 
All four algorithms guarantee system stability. 
Orth and rand-FGM tend to put the eigenvalues 
on the unit circle, yielding inaccurate system dynamics. 
SDP also has such a trend as $\bar T$ grows. 
However, \Sfv\ does not.}
\label{fig:polehis}
\end{center}
\end{figure}

\begin{figure}[t]
\begin{center}
\includegraphics[width=12cm]{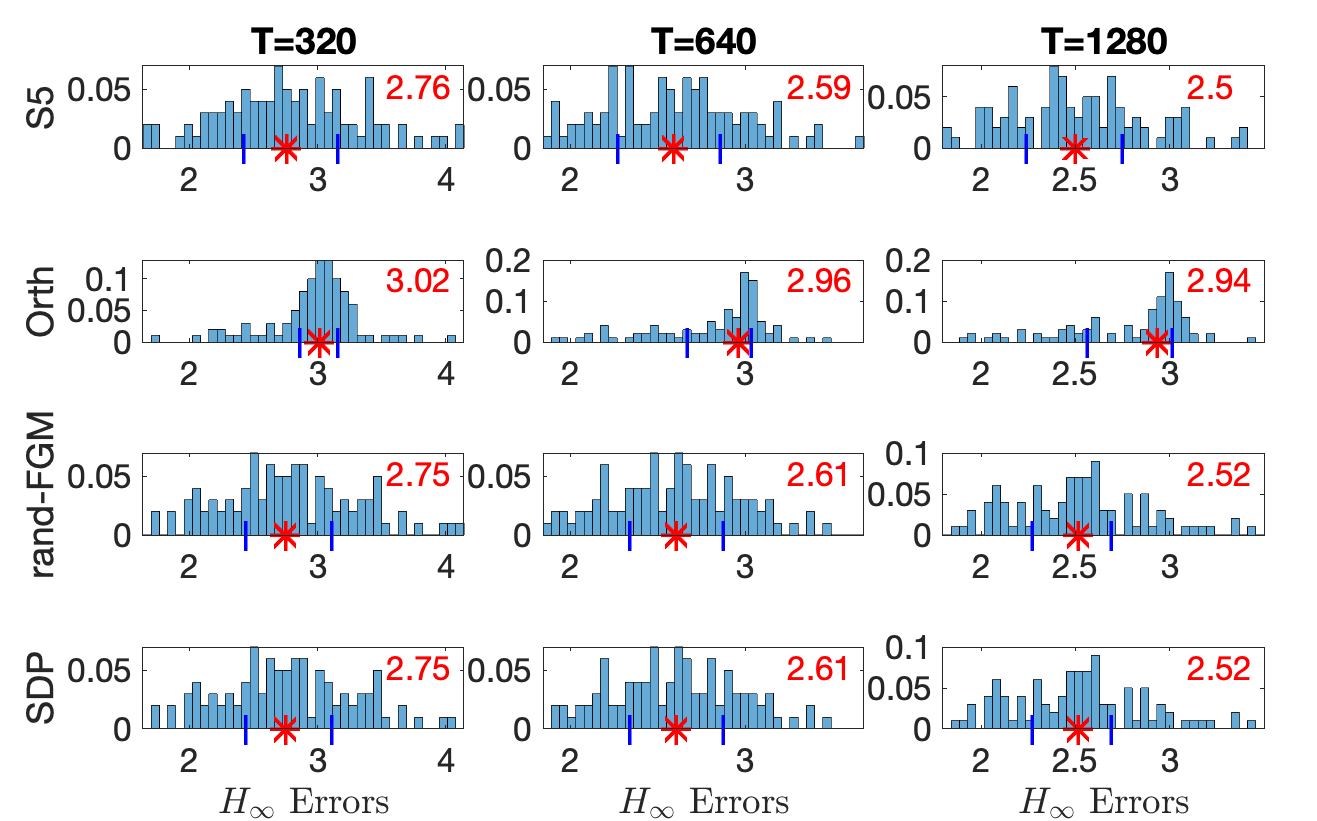}
\caption{Histograms of the soft $H_\infty$ error 
$e_3=\sup_{\omega\in[0,3]} \bar\sigma(\hat F(\omega)-F(\omega))$ for $n=5$: 
The medians are indicated by ``$*$'' marks and the number at the upper right corner 
of each histogram. The upper and lower quartiles are indicated by ``$|$'' marks.}
\label{fig:subHinf}
\end{center}
\end{figure}

\begin{figure}[t]
\hfill
\begin{minipage}[t]{1\linewidth}
  \centering
  \centerline{\includegraphics[width=10cm]{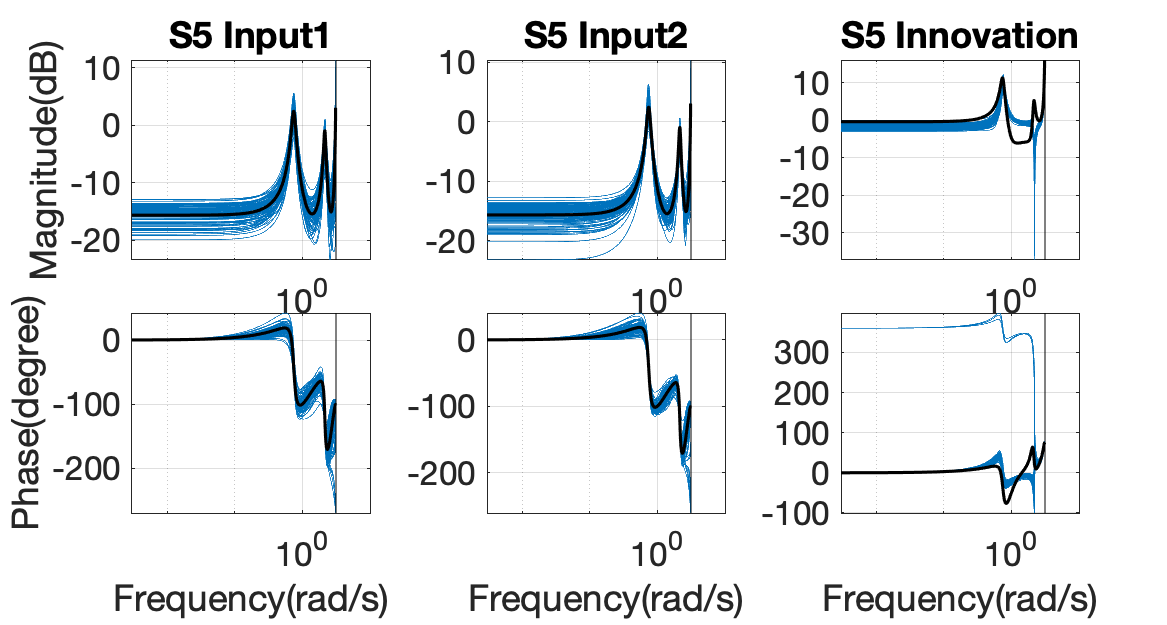} }
\end{minipage}
\begin{minipage}[t]{1\linewidth}
  \centering
  \centerline{\includegraphics[width=10cm]{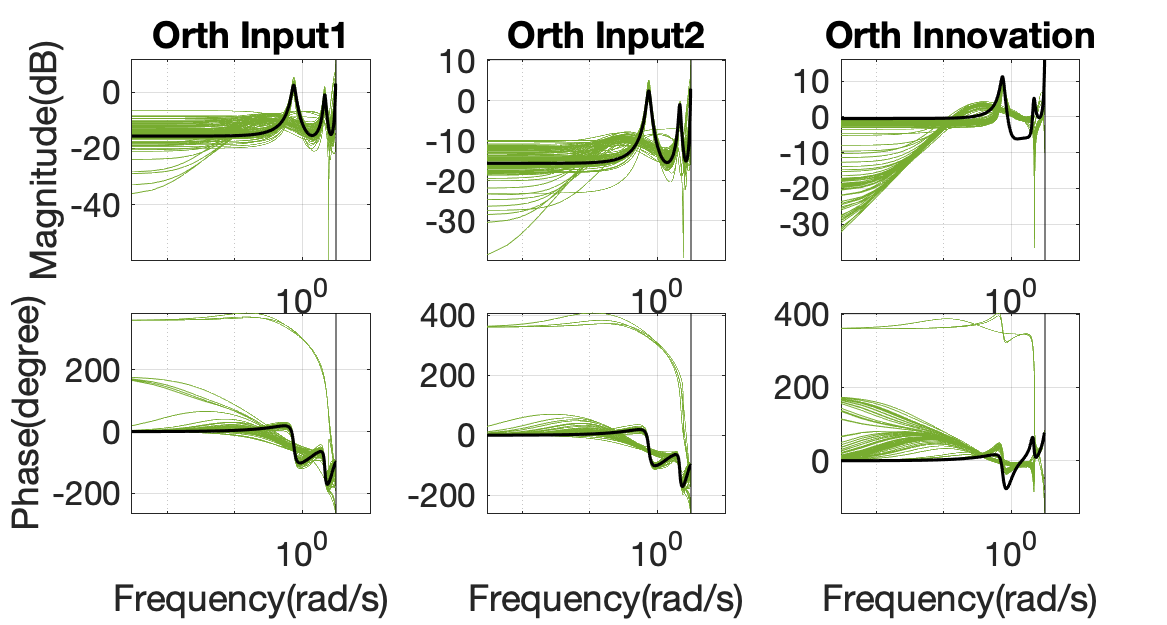} }
\end{minipage}
\caption{\label{fig:bode} $100$ estimated bode plots 
in frequency range $\omega\in[0,3]$ for $\bar T=1280$ for $n=5$: 
True bode plots are black. 
The bode plots for rand-FGM and SDP resemble closely to those of \Sfv\ 
and are thus omitted. 
In the undisplayed range $\omega\in(3,\pi]$, 
the peak magnitudes are around $45$dB, $300$dB, $200$dB and $110$dB 
for \Sfv, Orth, rand-FGM and SDP, respectively.}
\end{figure}

\subsection{Low Dimensional Simulations}
We first consider a low dimensional $2$-input-$1$-output ($m=2,d=1$) SS system 
with order $n=5$ and set the following true system matrices. 
\begin{align*}
A &= \left[\Bmat0.7& 0.642& 0& 0& 0\\
    -0.642& 0.7& 0& 0& 0\\
    0& 0& -0.5& 0.775& 0\\
    0& 0& -0.775& -0.5& 0\\
    0& 0& 0& 0& -0.995\Emat\right]\\
B &= \left[\Bsmat0.2& 0.2& 0.2& 0.2& 0.2\\
    0.2& 0.2& 0.2& 0.2& 0.2\Esmat\right]',\quad
C = \left[\Bsmat0.3& 0.3& 0.3& 0.3& 0.3\Esmat\right]\\
K &= \left[\Bmat0.5& 0.5& -0.3& -0.3& -0.9\Emat\right]', \quad 
\Qep = 1\\
A_u &= \left[\Bmat0.9&0.2\\-0.2&0.9\Emat\right], \quad 
Q_v = \lt[\Bmat1&0.5\\0.5&2\Emat\rt].
\end{align*}
System poles are $-0.995$, $0.95e^{\pm j0.74}$, $0.92e^{\pm j}$. 
$\bar A=A-KC$ has eigenvalues $0.686$, $0.803e^{\pm j0.917}$, $0.821e^{\pm j1.11}$.

We use the CVA subspace algorithm to get the state estimates $\hat X$. 
We set the future lag $f=10$ and vary $\bar T=320,640,1280$. 
We set the past lag $p = \lceil5\ln \bar T\rceil = 29,33,36$ for each $\bar T$. 

Since all algorithms use unstable LS estimates as initial values, 
we simulate a large number of repeats until there are $100$ unstable LS estimates 
for each $\bar T$. 
The empirical probabilities of having one unstable LS estimate 
are $5.24\%, 1.43\%, 0.069\%$ for $\bar T=320,640,1280$, respectively.

We identify the stable estimates for $A$ using the above $4$ algorithms. 
We implement Orth and FGM on SS identification by 
simply replacing the unstable LS estimates of $A$ by 
their `nearest' stable estimates. 
FGM is found to be most reliable when 
it tries random initial guesses of the stable matrix and picks the best one \cite{Nofe21}\cite{Gill19}. 
So we do the same and call it rand-FGM. 
MATLAB code for Orth and rand-FGM is available online \cite{Nofe-code}\cite{Gill-code}. 
Other tuning parameters, e.g. step sizes and convergence tolerances, of the Orth and rand-FGM 
are set as default. For SDP, we take the identity matrix as the weighting matrices and 
set the scale parameter to be $10^{-8}$.

\subsubsection{Inspection of Poles}
We first plot the pole locations of the estimated $A$ matrices 
in Fig. \ref{fig:poleloc} for complex poles 
and the pole histograms for the real dominating pole in Fig. \ref{fig:polehis}. 

From Fig. \ref{fig:poleloc}, we observe that 
expect for Orth, the estimated non-dominating poles 
cluster closely around the true ones. 
However, Orth prefers real poles which diverge far apart from the true poles. 
This is the same observation in their original paper \cite[Fig. 8]{Nofe21}. 

An inspection of Fig. \ref{fig:polehis} yields that 
all algorithms move the unstable LS poles inside the unit circle. 
However, Orth and rand-FGM tend to put poles very close to the unit circle. 
The percentages of poles with magnitudes greater than $1-10^{-10}$ 
for $\bar T=320,640,1280$ are: \\
$\bullet$ Orth: $\quad\quad\  87\%, 88\%, 85\%$, respectively. \\
$\bullet$ rand-FGM: $52\%, 62\%, 98\%$, respectively. \\
The above percentages are $0\%$ for \Sfv\ and SDP. 
However, SDP also has the same trend as $\bar T$ increases. 
A close inspection of \Sfv\ histograms shows 
that the largest magnitude of 
the estimated dominating pole is, 
on the contrary, decreasing as $\bar T$ increases, 
approaching closer to the true dominating pole. 
We need to emphasize that 
from the histograms, it may seem the \Sfv\ estimators 
are not statistically consistent, 
but that is because the plots are only for 
the small population 
($5.25\%$ for $\bar T=320$ and $0.069\%$ for $\bar T=1280$) 
of estimates where LS estimates are unstable.

\subsubsection{$H_\infty$ Errors}
We want to quantify the estimation error. 
Denoting by $\hat A_s$ any stable estimate from any algorithm, 
obvious choices are 
the perturbation mean squared error $\Vert \hat A_s-\hat A_*\Vert$ 
and the one-step-ahead prediction mean squared error 
$\Vert \hat X_1 - \hat A_s \hat X_0 - \hat B U_0\Vert$. 
However, the perturbation mean squared error has the described problems 
and the one-step-ahead prediction mean squared error 
does not fully depict the system dynamics; 
multi-step-ahead prediction error should also be considered. 
Thus, we consider the frequency behaviors and define the (hard) $H_\infty$ error
\EB
e = \sup_{\omega\in[0,\pi]} \bar\sigma(\hat F(\omega)-F(\omega)),
\end{align*}
where we have assumed the sampling frequency to be $1$ 
and $\bar\sigma(\cdot)$ denotes the largest singular value, 
$F(\omega) = C(e^{j\omega}I-A)^{-1}[B,K]+[0_{d\times m}, I_d]$ 
is the true frequency response of the whole SS system $(A,B,C,K)$, 
and $\hat F(\omega) = \hat C(e^{j\omega}I-\hat A_s)^{-1}[\hat B,\hat K]+[0_{d\times m}, I_d]$ 
is the estimated frequency response. 
However, inspecting the estimated poles 
suggests that when $\omega\approx\pi$, 
the hard $H_\infty$ error $e$ will be extremely large 
for Orth and rand-FGM since the estimated dominating poles are nearly on the unit circle, 
undermining good approximation at other frequencies. 
We thus also introduce the soft $H_\infty$ error
\EB
e_{\omega_*} = \sup_{\omega\in[0,\omega_*]} \bar\sigma(\hat F(\omega)-F(\omega)).
\end{align*}
We find taking $\omega_*=3$ best separates 
the well-approximated frequencies 
from those with an exploding response. 

We calculate both hard and soft $H_\infty$ 
errors by sampling the frequency responses 
on $1000$ points over the corresponding frequency range. 
We plot the histograms of the soft $H_\infty$ errors $e_3$ in Fig. \ref{fig:subHinf} 
and the estimated bode plots on top of the true ones in Fig. \ref{fig:bode}. 
We only plot the bode plots for $\bar T=1280$ due to space limits. 
{We also omit the bode plots for rand-FGM and SDP, 
since they resemble closely to those of \Sfv\ in the well-approximated frequency range. }
The hard $H_\infty$ errors differ greatly among the algorithms 
so they are summarized in Table. \ref{tbl:Hinf} as a better way of display. 

From Fig. \ref{fig:subHinf} and Fig. \ref{fig:bode}, it is observed that \Sfv, rand-FGM and SDP have 
nearly equally good performance in the well-approximated frequency region 
measured by the soft $H_\infty$ error whereas 
Orth shows a great discrepancy in the spectra 
resulting in the largest soft $H_\infty$ errors, 
because it tends to put the non-dominating poles on the real axis. 
For the response around the frequency $\pi$, 
inspection of Table. \ref{tbl:Hinf} yields clearly that 
\Sfv\ performs the best yielding magnitudes smaller 
hard $H_\infty$ errors than the other methods. 
Also, the increasing hard $H_\infty$ errors of SDP as $\bar T$ grows 
confirms the trend of its estimated dominating pole moving 
closer to the unit circle whereas our \Sfv\ has decreasing errors. 
Interestingly, our \Sfv\ estimates are by no means the `nearest' 
to the LS estimate but nonetheless have the best performance.

\subsubsection{Computational times}
Computational times are summarized in Table. \ref{tbl:lowcompt}. 
Our \Sfv\ computes magnitudes faster than the iterative algorithms.

\begin{table}[t]
\begin{center}
\begin{tabular}{|c| c c c c |}
\hline
Medians & S$^5$ & Orth & rand-FGM & SDP\\
\hline
$\bar T=320$ & $37.8$ & $7.2\times10^{14}$ & $2.7\times10^{9}$ & $90.6$\\
\hline
$\bar T=640$ & $34.6$ & $5.9\times10^{14}$ & $3.6\times10^{9}$ & $229$\\
\hline
$\bar T=1280$ & $28.6$ & $5.9\times10^{14}$ & $2.9\times10^{9}$ & $583$\\
\hline
\end{tabular}
\caption{\label{tbl:Hinf} Medians of the hard $H_\infty$ errors 
$e = \sup_{\omega\in[0,\pi]} \bar\sigma(\hat F(\omega)-F(\omega))$ for $n=5$.}
\end{center}
\end{table}

\begin{table}[t]
\begin{center}
\begin{tabular}{|c| c c c c |}
\hline
Total comp. time (s) & S$^5$ & Orth & rand-FGM & SDP\\
\hline
$\bar T=320$ & $0.0170$ & $5.40$ & $11.4$ & $1.26$\\
\hline
$\bar T=640$ & $0.0232$ & $5.27$ & $11.9$ & $1.30$\\
\hline
$\bar T=1280$ & $0.0340$ & $5.08$ & $15.9$ & $1.14$\\
\hline
\end{tabular}
\caption{\label{tbl:lowcompt} Total computational times (s) for 100 repeats 
 for $n=5$.}
\end{center}
\end{table}

\subsection{High Dimensional Simulations}
Aside from the superiority of \Sfv\ in 
both accuracy and computational efficiency, 
\Sfv\ scales easily to huge dimension identification. 
We illustrate this by considering model orders $n=16,64,256,1024$. 

However,  we only compare our \Sfv\ with rand-FGM here  
because in our previous work \cite{Rong23}, 
we showed that SDP can not handle system orders $n\geq250$,  
and in \cite{Nofe21}, it was noted that 
Orth can only handle system orders up to $n\approx100$. 

We design the true SS parameters as follows. 
We fix $d=1$ and $m=2$ and use the same $A_u$ and $Q_v$ 
as in the low-dimensional simulations. 
We put all system poles at the magnitude of $0.9999$ around the circle 
and set $A$ to be a block diagonal matrix. 
Each $2\times2$ block contains the real part of a pair of 
complex conjugate eigenvalues on its diagonal and 
the imaginary parts at its off-diagonal entries. 
We set all entries in $B$ to $0.2$ and those in $C$ to $0.3$. 
We choose the past and future lags $f=p=n+10$ 
and consider $\bar T=\lceil 5(m+2)f+500\rceil$. 
We set $K$ such that the the eigenvalues of $\bar A^p = (A-KC)^p$ 
have magnitude $0.1$. 
Such magnitude is neither too small so that the fit is too good to 
generate a single unstable LS estimate nor too big 
so that \Sfr\ generates bad estimates of the states 
(please refer to equation (\ref{eq:xt}) and the discussions below).
We simulate $50$ unstable LS estimates for each $n$. 
The empirical probabilities of getting one unstable LS estimate 
are $57.1\%, 94.3\%, 87.7\%, 3.1\%$ for $n=4,16,256,1024$, respectively. 

In the simulations, our \Sfv\ computed very fast with all model orders (see below). 
However, rand-FGM converged in all simulations 
for $n=16,64$ but it only converged in $26$ simulations for $n=256$ 
and in $37$ simulations for $n=1024$. 
The unconverged simulations resulted in NaN (Not a Number).

\begin{figure}[t]
\begin{center}
\includegraphics[width=12cm]{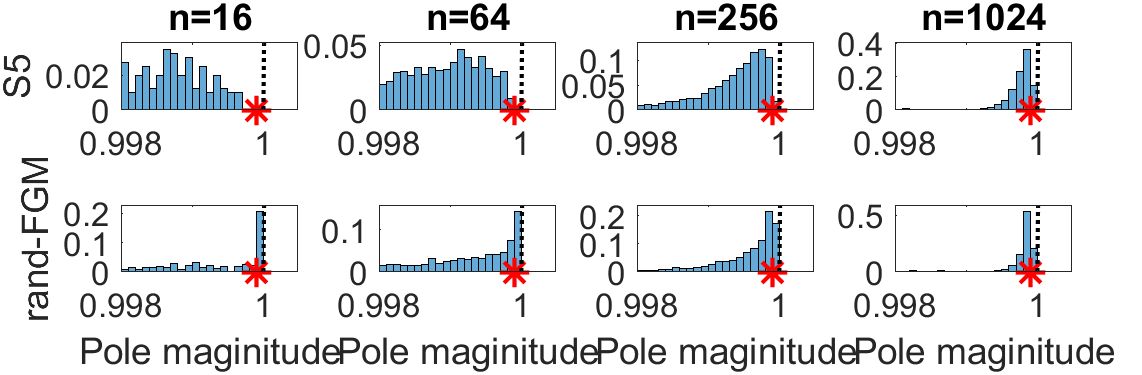}
\caption{Histograms of the pole magnitudes in high dimensional simulations: 
The ``$*$" marks are the true pole magnitude. 
The histograms for rand-FGM for $n=256$ and $n=1024$ 
use the $26$ and the $37$ converged estimates, respectively.}
\label{fig:polehish}
\end{center}
\end{figure}

\begin{figure}[t]
\begin{center}
\includegraphics[width=12cm]{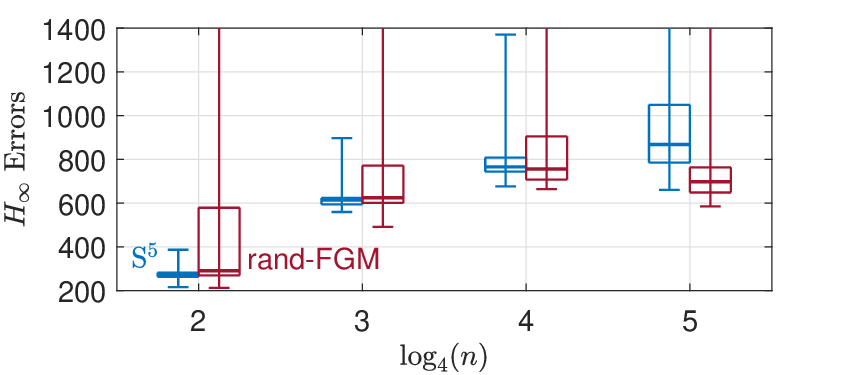}
\caption{Boxplots of $H_\infty$ errors in high-dimensional simulations: 
We only have $26$ and $37$ converged estimates with rand-FGM 
for $n=256$ and $n=1024$, respectively. 
The maximum $H_\infty$ errors for rand-FGM for $n=16,64,256,1024$ are, respectively, 
$7400, 90407, 3738,3481$ and that for \Sfv\ for $n=1024$ is $2417$; 
they are outside the range of the plot.}
\label{fig:Hinfh}
\end{center}
\end{figure}

We compare the converged estimates by 
\ben
\item in Fig. \ref{fig:polehish}, the pole magnitude histograms, 
\item in Fig. \ref{fig:Hinfh}, the (hard) $H_\infty$ error boxplots, and
\item in Table. \ref{tbl:compth}, the average computational times. 
\een
We have the following observations. 

Fig. \ref{fig:polehish} shows that the histograms 
of pole magnitudes of \Sfv\ cluster closer around the true pole magnitude 
as the order $n$ (as well as $\bar T$, noting $\bar T$ is a function of $n$) increases. 
However, rand-FGM still puts the poles very close to the unit circle 
for $n=16,64,256$. 

From Fig. \ref{fig:Hinfh}, \Sfv\ has much better $H_\infty$ errors for $n=16,64,256$
compared with the converged estimates of rand-FGM. 
Rand-FGM outperforms \Sfv\ in $H_\infty$ errors when $n=1024$ 
despite its poor convergence. 

From Table \ref{tbl:compth}, our \Sfv\ takes nearly no time to calculate 
even when $n=1024$, confirming the computational efficiency. 
However, rand-FGM takes hours for $n=1024$ with poor convergence.

\section{Conclusions}
In this paper, we considered a stable state space subspace (\Sfv) identification 
of an input-output state-space model with a VAR(1) input. 
We developed closed-form, stability-guaranteed estimators 
for both the state space model and the VAR(1) coefficient, 
based on correlation stable estimators. 
We showed that our estimators are strongly consistent 
under minor conditions. 

In comparative simulations, 
we considered both low- and high-dimensional cases 
with the model order up to $n=1024$. 
We compared our estimator 
with the existing iterative algorithms 
and demonstrated the superiority of \Sfv\ 
in both precision and computational cost. 

For future work, a natural extension is to consider a VAR(p) input. 
However, the consistency proof will be much much more complicated. 
We will pursue it elsewhere.

\begin{table}[t]
\begin{center}
\begin{tabular}{|c| c c  |}
\hline
Ave. comp. time& S$^5$ & rand-FGM\\
\hline
$n=16$ & $8.6\times10^{-4}$s & $0.17$s\\
\hline
$n=64$ & $4.7\times10^{-3}$s & $8.4$s\\
\hline
$n=256$ & $0.12$s & $402$s$=6.7$mins\\
\hline
$n=1024$ & $3.6$s & $8173$s$=2.3$hrs \\
\hline
\end{tabular}
\caption{\label{tbl:compth} Average computational times 
in high-dimensional simulations: 
For $n=256$ and $n=1024$, 
the average for rand-FGM is taken with the $26$ and the $37$ converged estimates.}
\end{center}
\end{table}

\bibliographystyle{plain}
\bibliography{S5ubib,sim}  
\end{document}